\documentstyle[12pt]{article}

\begin{document}

\title{On reliability function of quantum communication channel}
\author{M.~V.~Burnashev \thanks{%
Institute for Problems of Information Transmission, Russian Academy of
Sciences.}, A.~S.~Holevo\thanks{%
Steklov Mathematical Institute, Russian Academy of Sciences.}}
\date{}
\maketitle
\begin{center}
{\bf 1. Introduction}
\end{center}

Let ${\cal H}$ be a Hilbert space. We consider quantum channel \cite{hol73}
with a finite input alphabet $\{1,...,a\}$ and with pure signal states $S_i
= |\psi_i><\psi_i|; i=1,...,a$. Compound channel of length $n$ is in the $n$%
--tensor product of the space ${\cal H}\,$, i.e. $\,{\cal H}_{n} = {\cal H}
\otimes \ldots \otimes {\cal H}\,$. An input block (codeword) $\,u =
(i_{1},\ldots,i_{n})\,,\, i_{k} \in \{1,\ldots\,a\},\,$ for it means using
the compound state $\,\psi_{u} = \psi_{i_{1}} \otimes \ldots \otimes
\psi_{i_{n}} \in {\cal H}_{n}\,$ and the corresponding density operator $%
\,S_{u} = |\psi_{u}>< \psi_{u}|\,$ in the space $\,{\cal H}_{n}\,$. A code $%
\,({\cal C},{\bf X})\,$ of cardinality $\,M\,$ in $\,{\cal H}_n\,$ is a
collection of $\,M\,$ pairs $\,(u^{1}, X_{1}), \ldots, (u^{M}, X_{M}),\,$
where $\,{\bf X} = \{X_{1},\ldots,X_{M}, X_{M+1}\}\,$ is a quantum decision
rule, i.e. some resolution of identity in $\,{\cal H}_n\,$ \cite{hol73}. The
conditional probability $\,P(u^{i}|u^{k})\,$ to make a decision in favor of
message $\,u^{i}\,$ provided that message $\,u^{k}\,$ was transmitted is
given by 
\[
P(u^{i}|u^{k}) = {\rm Tr}\; S_{u^k}X_{i} = <\psi_{u^k}|X_{i}\psi_{u^k}>\;. 
\]
In particular, the probability to make a wrong decision when the message $%
\,u^k\,$ is transmitted is 
\[
1 - {\rm Tr}\; S_{u^k}X_{k} = 1 - <\psi_{u^k}|X_{k}\psi_{u^k}>\;. 
\]
For given code $\,({\cal C},{\bf X})\,$ of cardinality $\,M\,$ we consider
the following two error probabilities 
\[
\lambda_{max}({\cal C},{\bf X}) = \max_{1\leq k \leq M} [ 1 -
<\psi_{u^k}|X_{k}\psi_{u^k}>] \;, 
\]
\[
\bar{\lambda}({\cal C},{\bf X}) = {\frac{1 }{M}} \sum_{k=1}^{M} [1 -
<\psi_{u^k}|X_{k}\psi_{u^k}>]\;. 
\]
We shall denote by $P_e (M, n)$ any of the following two error probabilities 
\[
\lambda_{max}(M,n) = \inf_{{\cal C},{\bf X}}\, \lambda_{max} ({\cal C},{\bf X%
}),\qquad \bar{\lambda}(M,n) = \inf_{{\cal C},{\bf X}}\, {\bar \lambda} (%
{\cal C}, {\bf X}). 
\]
It is well--known in classical information theory \cite{gal} (Corollary 2 to
Theorem 5.6.2) that both $\lambda_{{\rm max}}(M,n)$ and $\bar{\lambda}(M,n)$
are essentially equivalent to each other. This remark obviously remains
valid for quantum channels as well.

The Shannon capacity of quantum channel was naturally defined in \cite{hol79}
as the number $C$ such that $P_e (\mbox{e}^{nR}, n)$ tends to zero as $%
n\rightarrow\infty$ for any $\,0 \leq R < C\,$ and does not tend to zero if $%
R>C$. Moreover, if $R<C$ then $P_e (\mbox{e}^{nR}, n)$ tends to zero
exponentially with $n$ and we are interested in the logarithmic rate of
convergence given by the reliability function 
\begin{equation}  \label{f1}
E(R) = \lim_{n \to \infty} \sup {\frac{1 }{n}}\ln {\frac{1 }{P_{e}(\mbox{e}%
^{nR},n)}} \;,\quad 0 < R < C \;.
\end{equation}
Our main results are the bounds for $E(R)$, reminiscent of the corresponding
bounds in the classical information theory \cite{gal}. It is also remarkable
that a number of tricks from the classical information theory works quite
well for quantum channels.

Let $\pi =\{\pi _{i}\}$ be a probability distribution on the input alphabet $%
\{1,...,a\}$, then we denote ${\bar{S}}_{\pi }=\sum_{i=1}^{a}\pi _{i}S_{i}$.
Let $\lambda _{j};j=1,...,d,$ where $d$ is the dimension of ${\cal H}$, be
the eigenvalues of operator ${\bar{S}}_{\pi }$ (obviously, forming a
probability distribution). Then 
\begin{equation}
H({\bar{S}}_{\pi })=-\mbox{Tr}{\bar{S}}_{\pi }\ln {\bar{S}}_{\pi
}=-\sum_{j=1}^{d}\lambda _{j}\ln \lambda _{j}  \label{f2}
\end{equation}
is the quantum entropy of the density operator ${\bar{S}}_{\pi }$. The upper
bound for the capacity 
\[
C\leq \max_{\pi }H({\bar{S}}_{\pi })
\]
follows directly from the {\sl entropy bound} \cite{hol73}. Recently, in 
\cite{jozsa} the converse inequality $C\geq \max_{\pi }H({\bar{S}}_{\pi })$
was established, implying the formula 
\begin{equation}
C=\max_{\pi }H({\bar{S}}_{\pi }).  \label{f3}
\end{equation}
(This result was generalized to arbitrary signal states in \cite{hol96}).
The proof in \cite{jozsa} was based on the notion of typical subspace
introduced in \cite{1schum}, \cite{schum}. A corollary of our estimates for
the reliability function is an alternative approach to the converse
inequality which makes no use of the notion of typical subspace.

\begin{center}
{\bf 2. The random coding lower bound}
\end{center}

Let $\bar{\lambda}(u^1,\ldots,u^M )$ be the average error probability
corresponding to codewords $u^1,...,u^M$ of the input alphabet of the length 
$n$ minimized over all quantum decision rules. Let $\pi = \{\pi_i\}$ be a
probability distribution on the input alphabet $\{1,...,a\}$ and assume that
the words are chosen at random, independently, and with the probability
distribution 
\begin{equation}  \label{f4}
{\sf P}\{u=(i_1 ,...,i_n )\} = \pi_{i_1}\cdot ...\cdot\pi_{i_n }
\end{equation}
for each word.

{\bf Proposition 1}. {\sl For all $M,n$ and $0\leq s\leq 1$ 
\begin{equation}
{\bf E}\,\bar{\lambda}(u^{1},\ldots ,u^{M})\leq 2(M-1)^{s}\left[ {\rm Tr}\,{%
\bar{S}}_{\pi }^{1+s}\right] ^{n}\,,  \label{f5}
\end{equation}
where} 
\[
{\rm Tr}\,{\bar{S}}_{\pi }^{1+s}=\sum_{j=1}^{d}\lambda _{j}^{1+s}\;. 
\]

{\sl Proof.}\footnote{The authors are grateful to T. Ogawa for pointing out an error in the earlier
version of the proof.} Let us for a while restrict to the subspace of ${\cal H}$
generated by the signal vectors $|\psi _{i}>;i=1,\ldots ,a$, and consider
the Gram matrix $\,\Gamma (u^{1},\ldots ,u^{M})=[<\psi _{u^{i}}|\psi
_{u^{j}}>]\,$ and the Gram operator $G(u^{1},\ldots
,u^{M})=\sum_{k=1}^{M}|\psi _{u^{k}}><\psi _{u^{k}}|$. This operator has the
matrix $\Gamma (u^{1},\ldots ,u^{M})$ in the (possibly overcomplete) basis 
\[
\hat{\psi}_{u^{i}}=G^{-1/2}(u^{1},\ldots ,u^{M}){\psi }_{u^{i}}\,;\,\quad
i=1,\ldots ,M\;. 
\]
In \cite{hol78} it was shown that using the resolution of identity of the
form $X_{k}=|\hat{\psi}_{k}><\hat{\psi}_{k}|$ we can upperbound the average
error probability as 
\begin{equation}
\bar{\lambda}(u^{1},\ldots ,u^{M})\leq {\frac{2}{M}}\,{\rm Sp}\left(
E-\Gamma ^{1/2}(u^{1},\ldots ,u^{M})\right) ,  \label{f6}
\end{equation}
where $E$ is the unit $M\times M$-matrix and ${\rm Sp}$ is the trace of $%
M\times M$-matrix. Indeed, 
\[
\bar{\lambda}(u^{1},\ldots ,u^{M})={\frac{1}{M}}\sum_{k=1}^{M}[1-|<\psi
_{u^{k}}|{\hat{\psi}}_{u^{k}}>|^{2}] 
\]
\[
\leq {\frac{2}{M}}\sum_{k=1}^{M}[1-|<\psi _{u^{k}}|{\hat{\psi}}_{u^{k}}>|]={%
\frac{2}{M}}\sum_{k=1}^{M}[1-<{\hat{\psi}_{u^{k}}}|G^{1/2%
}(u^{1},\ldots ,u^{M}){\hat{\psi}}_{u^{k}}>], 
\]
which is (6).

The first step of our argument is to remark that

\begin{equation}  \label{f7}
{\frac{2}{M}}\,{\rm Sp(}E-\Gamma ^{1/2}(u^{1},\ldots ,u^{M}))={\frac{2}{M}}%
(M-{\rm Tr}G^{1/2}(u^{1},\ldots ,u^{M})).
\end{equation}
In what follows we shall skip $u^{1},\ldots ,u^{M}$ from notations. Consider
two operator inequalities 
\[
-2G^{1/2}\leq -2G+2G, 
\]
\[
-2G^{1/2}\leq -2G+(G^{2}-G). 
\]
The first one is obvious, while the second follows from the inequality 
\[
-2x^{1/2}=2(1-x^{1/2})-2=(1-x^{1/2})^{2}-1-x\leq
(1-x)^{2}-1-x=x^{2}-3x, 
\]
valid for $x\geq 0.$ Taking the expectation with respect to the probability
distribution (\ref{f3}), we get 
\[
-2\,{\bf E}G^{1/2}\leq -2{\bf E}G+\left\{ 
\begin{array}{l}
2\,{\bf E}G \\ 
{\bf E}(G^{2}-G)
\end{array}
\right. . 
\]
Now 
\[
{\bf E}G={\bf E}\sum_{k=1}^{M}|\psi _{u^{k}}><\psi _{u^{k}}|=M{\bf E}|\psi
_{u^{k}}><\psi _{u^{k}}|=M{\bar{S}}_{\pi }^{\otimes n}, 
\]
\[
{\bf E}(G^{2}-G)={\bf E}\sum_{k=1}^{M}\sum_{l=1}^{M}|\psi _{u^{k}}><\psi
_{u^{k}}||\psi _{u^{l}}><\psi _{u^{l}}|-{\bf E}\sum_{k=1}^{M}|\psi
_{u^{k}}><\psi _{u^{k}}|
\]
\[
={\bf E}\sum_{k\neq l}|\psi _{u^{k}}><\psi _{u^{k}}|\psi _{u^{l}}><\psi
_{u^{l}}| =M(M-1)\left[ {\bar{S}}_{\pi }^{\otimes n}\right] ^{2}. 
\]
Let$\left\{ e_{J}\right\} $ be the orthonormal basis of eigenvectors, and $%
\lambda _{J}$ the corresponding eigenvalues of the operator ${\bar{S}}_{\pi
}^{\otimes n}.$ Then 
\[
-2\left\langle e_{J}|{\bf E}G^{1/2}|e_{J}\right\rangle \leq
= -2M\lambda _{J}+M\lambda _{J}\min \left( 2,(M-1)\lambda _{J}\right) . 
\]
Using the inequality $\,\min \{a,b\}\leq a^{s}b^{1-s},\,0\leq s\leq 1,$ we
get 
\[
\min \left( 2,(M-1)\lambda _{J}\right) \leq 2(M-1)^{s}\lambda
_{J}^{1+s}\,,\;0\leq s\leq 1\,. 
\]
Summing with respect to $J$ and dividing by $M,$ we get from (\ref{f6}), (%
\ref{f7}) 
\[
{\bf E}\,\bar{\lambda}(u^{1},\ldots ,u^{M})\leq 2(M-1)^{s}\sum_{J}\lambda
_{J}^{1+s}=2(M-1)^{s}\left[ {\rm Tr}\,{\bar{S}}_{\pi }^{1+s}\right]
^{n}\,,\;0\leq s\leq 1\,.\qquad \Box 
\]

It is natural to introduce the function $\mu(\pi,s)$ similar to analogous
function in classical information theory (e.g. \cite{gal}, Ch. 5) 
\begin{equation}  \label{f19}
\mu (\pi,s) = - \ln {\rm Tr}\,{\bar S}_{\pi}^{1+s} = - \ln \sum_{j=1}^d
\lambda_j^{1+s}.
\end{equation}
Clearly, $\mu(\pi,0) = 0$. Using the formulas 
\[
\mu^{\prime}(\pi,s) = -\frac{{\rm Tr}\,{\bar S}_{\pi}^{1+s}\ln {\bar S}_{\pi}%
} {{\rm Tr}\,{\bar S}_{\pi}^{1+s}}, 
\]
\[
\mu^{\prime\prime}(\pi,s) = \frac{({\rm Tr}\,{\bar S}_{\pi}^{1+s} \ln {\bar S%
}_{\pi})^2 - {\rm Tr}\,{\bar S}_{\pi}^{1+s} (\ln {\bar S}_{\pi})^2 {\rm Tr}\,%
{\bar S}_{\pi}^{1+s}} {({\rm Tr}\,{\bar S}_{\pi}^{1+s})^2}\,, 
\]
it is easy to check that $\mu(\pi,s)$ is nondecreasing and $\cap$--convex
function of $s$. Moreover, 
\[
\mu^{\prime}(\pi,0)=H({\bar S}_{\pi})\,. 
\]

There is special case where among the signal vectors $|\psi_i\rangle ; i=1,
\dots , a$, there are $k\geq 1$ mutually orthogonal: then defining $\pi$
to be the uniform distribution on these vectors, one has $\mu (\pi,
s) = s\ln k$. Otherwise the function $\mu (\pi, s)$ is strictly increasing
and strictly $\cap$--convex (see Fig. 1).

By taking $M= \mbox{e}^{nR}$, we obtain

{\bf Corollary 1.} {\it For any $0 < R < C$ the following lower bound holds} 
\begin{equation}  \label{f20}
E(R) \geq \max_{\pi} \max_{0\leq s \leq 1} \left(\mu(\pi,s)-s R\right)\,
\equiv E_r (R).
\end{equation}

{\bf Corollary 2.} 
\begin{equation}  \label{f21}
C \geq \max_{\pi}H({\bar S}_{\pi}).
\end{equation}

{\sl Proof}. Indeed, 
\[
C \geq \max_{\pi} \max_{0\leq s \leq 1} {\frac{\mu(\pi,s) }{s}} \geq
\max_{\pi} \mu^{\prime}(\pi,0) = \max_{\pi}H({\bar S}_{\pi})\,. \;\Box. 
\]

Inequality (\ref{f21}) together with the converse inequality from \cite
{hol73} provides an alternative proof of the formula (\ref{f3}).

The maximization with respect to $s$ can be treated in the same way as in
Ch. 5.7 of \cite{gal}. Defining the function 
\begin{equation}  \label{f22a}
E_r (\pi , R) = \max_{0\leq s \leq 1}[ \mu (\pi , s) - sR],
\end{equation}
we have 
\[
E_r (\pi, R) = \mu (\pi, 1) - R \quad\mbox{for}\quad 0\leq R\leq
\mu^{\prime}(\pi, 1), 
\]
where 
\[
\mu (\pi, 1) = - \ln \mbox{Tr}{\bar S}_{\pi}^2 = - \ln \sum_{i,j =1}^a \pi_i
\pi_j |<\psi_i |\psi_j >|^2. 
\]
For $\mu^{\prime}(\pi, 1)\leq R < C$ the function $E_r (\pi, R)$ is a $\cup$%
--convex and is given by 
\[
E_r (\pi, R) = \mu(\pi, s_R)-s_R R, 
\]
where $s_R$ is the root of the equation $\mu^{\prime}(\pi, s_R)=R$ (see Fig.
2). 

\begin{center}
{\bf 3. The expurgated lower bound}
\end{center}

When we choose codewords randomly there is certain probability that some
codewords will coincide that makes error probability for such code equal to $%
1$. It turns out that probability to choose such a bad code does not
influence essentially the average code error probability if the rate $R$ is
relatively high. Conversely, it becomes dominating for low rates $R$. All
these effects are well described in \cite{gal}, Ch. 5.7. In order to reduce
the influence of choosing such bad codes an elegant "expurgation" technique
has been developed.

We start with an ensemble of codes with $M^{\prime}= 2M - 1$ codewords and
denote by 
\[
\lambda_k = [1 - |<\psi_{u^k}|{\hat \psi}_{u^k}>|^2 ] 
\]
the probability of erroneous decision for the word $u^k$, when the decision
rule \{$X_k$\} from Sec. 2 is used. Then according to the Lemma from Ch. 5.7 
\cite{gal} there exists a code in the ensemble of codes with $M^{\prime}= 2M
- 1$ codewords, for which at least $M$ codewords satisfy 
\begin{equation}  \label{f24}
\lambda_k \leq \left[2 {\bf E} {\lambda}_k^r \right]^{1/r} \;,
\end{equation}
for arbitrary $0< r\leq 1$ (without loss of generality we can assume that (%
\ref{f24}) holds for $k = 1, \ldots, M$). Then we can use an estimate from 
\cite{jozsa} to evaluate the righthand side of (\ref{f24}). By using the
inequality $\sqrt{\gamma} \geq {\frac{3 }{2}}\gamma - {\frac{1 }{2}}\gamma^2$
for $\gamma\geq 0$, one obtains 
\[
\lambda_k\leq 2[1 - <{\hat \psi}_{u^k}|G^{\frac{1 }{2}}(u^1, \ldots,
u^{M^{\prime}}) {\hat \psi}_{u_k}>] 
\]
\[
\leq 2 - 3<{\hat \psi}_{u^k}|G (u^1, \ldots, u^{M^{\prime}}) {\hat \psi}%
_{u_k}> + <{\hat \psi}_{u^k}|G^2 (u^1, \ldots, u^{M^{\prime}}) {\hat \psi}%
_{u_k}> 
\]
\[
= \sum_{i\not{= }k} |<\psi_{u^i}|\psi_{u^k}>|^2, 
\]
where the summation is over $i$ from 1 to $M^{\prime}$.

Using the inequality $\,(\sum a_{i})^r \leq \sum a_{i}^r\,,\, 0 < r \leq 1,
\,$ we get for randomly chosen codewords 
\[
{\bf E}\lambda_k^r \leq (M^{\prime}- 1){\bf E}\,
|<\psi_{u}|\psi_{u^{\prime}}>|^{2r} = 2 (M -1 )\left[\sum_{i,k = 1}^{a}\pi_i
\pi_k |<\psi_i |\psi_k >|^{2r}\right]^n . 
\]
Substituting this into (\ref{f24}) and denoting $s = {\frac{1 }{r}}$, we
obtain

{\bf Proposition 2}. {\sl For all} $s\geq 1$ 
\begin{equation}  \label{f25}
\lambda_{max}(M, n)\leq \left( 4(M - 1)\left[ \sum_{i,k = 1}^{a}\pi_i \pi_k
|<\psi_i |\psi_k>|^{\frac{2 }{s}}\right] ^n \right) ^{s}.
\end{equation}

Taking $M = \mbox{e}^{nR}$, we obtain the lower bound with expurgation 
\[
E(R)\geq \max_{\pi}\max_{s \geq 1} ({\tilde \mu}(\pi, s) - s (R + {\frac{\ln
4 }{n}}))\equiv E_{ex}(R +{\frac{\ln 4 }{n}}), 
\]
where 
\[
{\tilde \mu}(\pi, s) = - s \ln \sum_{i,k = 1}^{a}\pi_i \pi_k |<\psi_i
|\psi_k>|^{\frac{2 }{s}}. 
\]

The function ${\tilde \mu}(\pi, s)$ is $\cap$-convex, increasing from the
value 
\[
{\tilde \mu}(\pi, 1) = \mu (\pi, 1) = - \ln\mbox{Tr}{\bar S}_{\pi}^2 
\]
for $s = 1$ to 
\[
{\tilde \mu}(\pi, \infty) = - \sum_{i,k=1}^a \pi_i \pi_k \ln |<\psi_i
|\psi_k >|^2, 
\]
(which may be infinite if there are orthogonal states). The behavior of ${%
\tilde \mu}(\pi, s)$ in the case where this value is finite is shown on Fig.
1.

By introducing 
\begin{equation}  \label{f24a}
E_{ex} (\pi, R) = \max_{s\geq 1}[{\tilde \mu}(\pi, s) -sR],
\end{equation}
we can investigate the behavior of $E_{ex}(\pi, R)$ like in the classical
case. Namely, for $0< R\leq {\tilde \mu}^{\prime}(\pi, 1)$, where 
\[
{\tilde \mu}^{\prime}(\pi, 1) = -\ln\mbox{Tr}{\bar S}_{\pi}^2 + {\frac{{\
\sum_{i,k=1}^a \pi_i \pi_k |<\psi_i |\psi_k >|^2 \ln|<\psi_i |\psi_k >|^2} }{%
{\mbox{Tr}{\bar S}_{\pi}^2}}} \leq {\tilde \mu}(\pi, 1), 
\]
the function $E_{ex}(\pi, R)$ is $\cup$-convex decreasing from 
\begin{equation}  \label{f250}
E_{ex}(\pi, +0) = {\tilde \mu}(\pi, \infty ) = - \sum_{i,k=1}^a \pi_i \pi_k
\ln |<\psi_i |\psi_k >|^2
\end{equation}
to $E_{ex}( {\tilde \mu}^{\prime}(\pi, 1)) = {\tilde \mu}(\pi, 1) - {\tilde
\mu}^{\prime}(\pi, 1)$. In the interval ${\tilde \mu}^{\prime}(\pi, 1) \leq
R \leq {\tilde \mu}(\pi, 1)$ it is linear function 
\[
E_{ex}(\pi, R) = {\tilde \mu}(\pi, 1) - R, 
\]
and $E_{ex}(\pi, R) = 0$ for ${\tilde \mu}(\pi, 1)\leq R < C$.

Thus comparing it with $E_r (\pi, R)$ we have in generic case the picture on
Fig. 2: 
\[
\begin{array}{ll}
E_r(\pi, R) < E_{ex}(\pi, R),\qquad & 0 < R \leq {\tilde \mu}^{\prime}(\pi,
1); \\ 
E_r(\pi, R) = E_{ex}(\pi, R), \qquad & {\tilde \mu}^{\prime}(\pi, 1)\leq R <
\mu ^{\prime}(\pi, 1); \\ 
E_r(\pi, R) > E_{ex}(\pi, R),\qquad & \mu ^{\prime}(\pi, 1)\leq R <C.
\end{array}
\]
However, it may happen that ${\tilde \mu}^{\prime}(\pi, 1) > \mu
^{\prime}(\pi, 1)$, in which case the linear piece of the bound is absent.

The value (\ref{f250}) is in fact exact as the following proposition shows.

{\bf Proposition 3.} {\it If $\,|<\psi_{i}|\psi_{k}>| > 0\,$ for any $\,1
\leq i,k \leq a\,$ then 
\begin{equation}  \label{f26}
E(+0) = - \min_{\{\pi\}}\; \sum_{i,k=1}^{a} \pi_{i}\pi_{k}\ln
|<\psi_{i}|\psi_{k}>|^2 .
\end{equation}
}

{\it If $\,|<\psi_{i}|\psi_{k}>| = 0\,$ for some $\,i,k,\,$ then $\,E(+0) =
\infty\,$. }

{\it Proof.} From (\ref{f250}) we see that $E (+0)$ is greater than or equal
to the righthand side of (\ref{f26}). On the other hand, from testing two
hypotheses (\cite{hel}, p. 130, (2.34)), we have 
\[
\lambda_{max} ({\cal C}, {\bf X}) \geq {\frac{1 }{2}} \left[1 - \sqrt{1 -
\max_{u \neq u^{\prime}} |<\psi_{u}|\psi_{u^{\prime}}>|^{2}} \right] \geq {%
\frac{1 }{4}} \max_{u \neq u^{\prime}} |<\psi_{u}|\psi_{u^{\prime}}>|^{2}
\;, 
\]
where $u, u^{\prime}$ are codewords from ${\cal C} = (u^1, \ldots, u^M )$,
and therefore 
\[
E(+0) \leq - \lim_{n \to \infty} \left[ {\frac{2 }{n}}\, \max_{u \neq
u^{\prime}}\;\ln |<\psi_{u}|\psi_{u^{\prime}}>|\right] \;. 
\]
Denote $\,\psi_{u}(k)\,$ the $k$--th component of the codeblock $u$ and let $%
\,k_{i} = \pi_{i}M\,$ be the number of codeblocks $\,u\,$ with $%
\,\psi_{u}(k)=\psi_{i}\,,\,i=1,\ldots,d\,$. Then we have 
\[
\max_{u \neq u^{\prime}}\;\ln |<\psi_{u}|\psi_{u^{\prime}}>| \geq {\frac{1 }{%
M(M-1)}} \sum_{u, u^{\prime}\in {\cal C}} \ln |<\psi_{u}|\psi_{u^{\prime}}>| 
\]
\[
\geq {\frac{n }{M(M-1)}} \min_{1\leq k \leq n} \; \sum_{u, u^{\prime}\in%
{\cal C}} \ln |<\psi_{u}(k)|\psi_{u^{\prime}}(k)>| 
\]
\[
\geq {\frac{n }{M(M-1)}} \min_{\{k_{i}\}} \left\{
\sum_{i=1}^{a}\sum_{j=1}^{a} k_{i}k_{j} \ln |<\psi_{i}|\psi_{j}>| \right\} 
\]
\[
\geq {\frac{nM }{(M-1)}} \min_{\{\pi\}} \left\{ \sum_{i=1}^{a}\sum_{j=1}^{a}
\pi_{i}\pi_{j} \ln |<\psi_{i}|\psi_{j}>| \right\} \;, 
\]
from where it follows 
\[
E(+0) \leq - \min_{\{\pi\}}\; \sum_{i,k=1}^{a} \pi_{i}\pi_{k}\ln
|<\psi_{i}|\psi_{k}>|^2 . 
\]
In a result, we get (\ref{f26}).$\Box$

\begin{center}
{\bf 4. The binary quantum channel}
\end{center}

Maximization of the bounds $E_{r}(\pi, R), E_{ ex}(\pi, R)$ over $\pi$,
which is a difficult problem even in the classical case, is still more
difficult in quantum case. However, if the distribution ${\pi}^0$ maximizing
either $\mu (\pi, s)$ or ${\tilde \mu}(\pi, s)$ is the same for all $s$,
then the analysis of Secs. 2, 3 applies to functions 
\[
E_r (R) = E_r ({\pi}^0, R),\qquad E_{ex} (R) = E_{ ex} ({\pi}^0, R). 
\]

Let $a=d=2$ and $\,|\psi_0>,\, |\psi_1>$ be two pure states with $%
|<\psi_0|\psi_1>| = \epsilon $. Consider the operator $\,S_{\pi} = (1-\pi)
S_{0} + \pi S_{1}\,$. Its eigenvectors have the form $|\psi_0> + \alpha
|\psi_1>$ with some $\alpha$. Therefore for its eigenvalues we get the
equation 
\[
\left((1-\pi) |\psi_0><\psi_0| + \pi |\psi_1><\psi_1|\right) \left(|\psi_0>
+ \alpha |\psi_1>\right) = \lambda \left(|\psi_0> + \alpha |\psi_1>\right). 
\]
Solving it, we find the eigenvalues 
\[
\lambda_{1}(\pi) = {\frac{1 }{2}} \left[1 - \sqrt{1 - 4(1-\epsilon^{2})\pi
(1-\pi)}\right] \;, 
\]
\[
\lambda_{2}(\pi) = {\frac{1 }{2}} \left[1 + \sqrt{1 - 4(1-\epsilon^{2})\pi
(1-\pi)}\right] \;. 
\]

It is easy to check that both functions 
\[
\mu (\pi, s) = - \ln \left(\lambda_1 (\pi )^{1+s} + \lambda_2 (\pi
)^{1+s}\right), 
\]
\[
{\tilde \mu}(\pi, s) = - s \ln\left( \pi^2 + (1-\pi )^2 + 2\pi (1 - \pi
)\epsilon^{2/s}\right) 
\]
are maximized by $\pi = 1/2$. Denoting 
\[
\mu (s) = \mu(1/2, s) = - \ln \left[\left({\frac{1-\epsilon }{2}}%
\right)^{1+s} + \left({\frac{1+\epsilon }{2}}\right)^{1+s}\right], 
\]
\[
{\tilde \mu}(s) = {\tilde \mu}(1/2, s) = - s \ln\left[{\frac{1+{\epsilon}%
^{2/s} }{2}}\right], 
\]
we get the following bound 
\[
\begin{array}{ll}
E (R)\geq {\tilde \mu}({\tilde s}_R) - {\tilde s}_R R,\quad & 0< R\leq {%
\tilde \mu}^{\prime}(1); \\ 
E (R)\geq \mu (1) - R, \quad & {\tilde \mu}^{\prime}(1)\leq R\leq {\mu}%
^{\prime}(1); \\ 
E (R)\geq \mu(s_R ) - s_R R, \quad & {\mu}^{\prime}(1)\leq R <C,
\end{array}
\]
where ${\tilde s}_R, s_R$ are solutions of the equations ${\tilde \mu}%
^{\prime}({\tilde s}_R) = R,\quad {\mu}^{\prime}(s_R) = R,$ 
\[
\mu (1) = {\tilde \mu} (1) = - \ln \left({\frac{1+\epsilon^2 }{2}} \right), 
\quad
{\tilde \mu}^{\prime}(1) = {\tilde \mu}(1) + {\frac{\epsilon^2\ln\epsilon^2 
}{1+\epsilon^2}}, 
\]
\[
{\mu}^{\prime}(1) = - {\frac{(1-\epsilon)^{2}\ln \left({\frac{1-\epsilon }{2}%
}\right) + (1+\epsilon)^{2}\ln \left({\frac{1+\epsilon }{2}}\right) }{%
2(1+\epsilon^{2})}} \,, 
\]
\[
C = {\mu}^{\prime}(0) = - \left[\left({\frac{1-\epsilon }{2}}\right) \ln
\left({\frac{1-\epsilon }{2}}\right) + \left({\frac{1+\epsilon }{2}}\right)
\ln \left({\frac{1+\epsilon }{2}}\right)\right]. 
\]
Moreover, from Proposition 3, 
\[
E(+0)= -\ln\epsilon. 
\]

\begin{center}
{\bf 5. Comments on the case of arbitrary signal states}
\end{center}

The case where the signal states are given by {\sl commuting} density
operators $S_{i}$ reduces to the case of classical channel with transition
probabilities $\lambda _{j}^{i}$, where $\lambda _{j}^{i}$ are the
eigenvalues of $S_{i}$. The classical bound given by Theorem 5.6.1 \cite{gal}
in the case of commuting density operators takes the form 
\begin{equation}
{\bf E}\bar{\lambda}(u^{1},\ldots ,u^{M})\leq \min_{0<s\leq
1}(M-1)^{s}\left( \mbox{Tr}\left[ \sum_{i=1}^{a}\pi _{i}S_{i}^{\frac{1}{1+s}%
}\right] ^{1+s}\right) ^{n}.  \label{f27}
\end{equation}

The expurgated bound given by Theorem 5.7.1 of \cite{gal} in the case of
commuting density operators reads 
\begin{equation}
\lambda _{max}(M,n)\leq \min_{s\geq 1}\left( 4(M-1)\left[
\sum_{i,k=1}^{a}\pi _{i}\pi _{k}(\mbox{Tr}\,\sqrt{S_{i}}\sqrt{S_{k}})^{%
\frac{1}{s}}\right] ^{n}\right) ^{s}.  \label{f28}
\end{equation}
The righthand sides of (\ref{f27}), (\ref{f28}) are meaningful for arbitrary
density operators, which gives some hope that these estimates could be
generalized to the noncommutative case with minor modifications. \vskip10pt 
{\sl Acknowledgements.} This work was partially supported by RFBR grants
no.95-01-00136 and 96-01-01709. The second author is grateful to
participants of Prof. M. S. Pinsker's seminar in the Institute for Problems
of Information Transmission for enlightening discussion, which in particular
stimulated search of the proof of quantum coding theorem, independent of the
notion of typical subspace. He acknowledges also the hospitality of the
Electrical and Computer Engineering Department of the Northwestern
University, where the improved version of the paper was written.

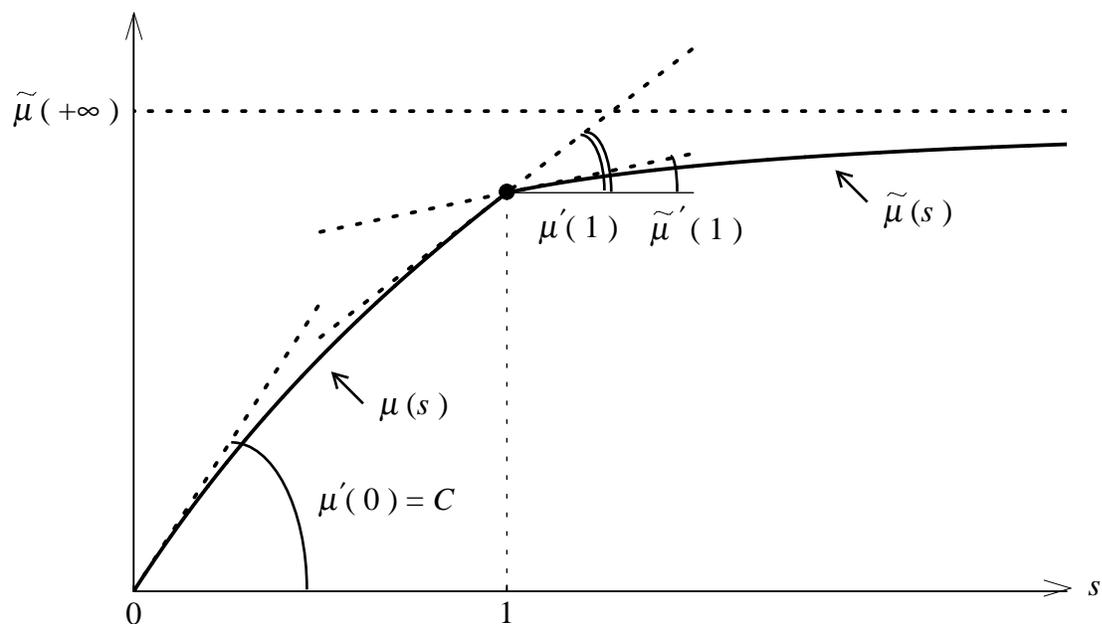
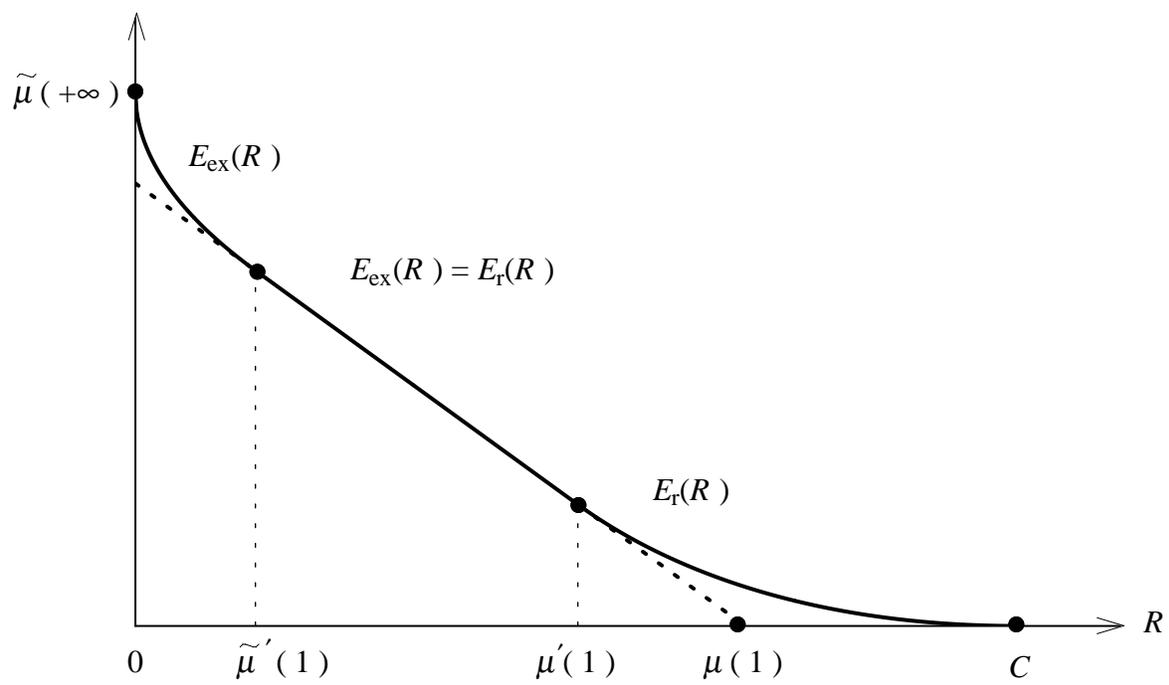
\end{document}